\documentclass[modern]{aastex61}

\newcommand\txs{TXS 0506+056}
\newcommand\icnu{IC\,170922A}

\submitjournal{ApJL}

\shorttitle{VERITAS observations of TXS 0506+056}
\shortauthors{Abeysekara et al.}

\begin{document}

\title{VERITAS observations of the BL Lac object TXS 0506+056}

\author{ A.~U.~Abeysekara }
\affiliation{ Department of Physics and Astronomy, University of Utah, Salt Lake City, UT 84112, USA }

\author{ A.~Archer }
\affiliation{ Department of Physics, Washington University, St. Louis, MO 63130, USA }

\author{ W.~Benbow }
\affiliation{ Fred Lawrence Whipple Observatory, Harvard-Smithsonian Center for Astrophysics, Amado, AZ 85645, USA }

\author{ R.~Bird }
\affiliation{ Department of Physics and Astronomy, University of California, Los Angeles, CA 90095, USA }

\author{ A.~Brill }
\affiliation{ Physics Department, Columbia University, New York, NY 10027, USA }

\author{ R.~Brose }
\affiliation{ Institute of Physics and Astronomy, University of Potsdam, 14476 Potsdam-Golm, Germany }
\affiliation{ DESY, Platanenallee 6, 15738 Zeuthen, Germany }

\author{ J.~H.~Buckley }
\affiliation{ Department of Physics, Washington University, St. Louis, MO 63130, USA }

\author{ J.~L.~Christiansen }
\affiliation{ Physics Department, California Polytechnic State University, San Luis Obispo, CA 94307, USA }

\author{ A.~J.~Chromey }
\affiliation{ Department of Physics and Astronomy, Iowa State University, Ames, IA 50011, USA }

\author{ M.~K.~Daniel }
\affiliation{ Fred Lawrence Whipple Observatory, Harvard-Smithsonian Center for Astrophysics, Amado, AZ 85645, USA }

\author{ A.~Falcone }
\affiliation{ Department of Astronomy and Astrophysics, 525 Davey Lab, Pennsylvania State University, University Park, PA 16802, USA }

\author{ Q.~Feng }
\affiliation{ Physics Department, McGill University, Montreal, QC H3A 2T8, Canada }

\author{ J.~P.~Finley }
\affiliation{ Department of Physics and Astronomy, Purdue University, West Lafayette, IN 47907, USA }

\author{ L.~Fortson }
\affiliation{ School of Physics and Astronomy, University of Minnesota, Minneapolis, MN 55455, USA }

\author{ A.~Furniss }
\affiliation{ Department of Physics, California State University - East Bay, Hayward, CA 94542, USA }

\author{ G.~H.~Gillanders }
\affiliation{ School of Physics, National University of Ireland Galway, University Road, Galway, Ireland }

\author{ O.~Gueta }
\affiliation{ DESY, Platanenallee 6, 15738 Zeuthen, Germany }

\author{ D.~Hanna }
\affiliation{ Physics Department, McGill University, Montreal, QC H3A 2T8, Canada }

\author{ O.~Hervet }
\affiliation{ Santa Cruz Institute for Particle Physics and Department of Physics, University of California, Santa Cruz, CA 95064, USA }

\author{ J.~Holder }
\affiliation{ Department of Physics and Astronomy and the Bartol Research Institute, University of Delaware, Newark, DE 19716, USA }

\author{ G.~Hughes }
\affiliation{ Fred Lawrence Whipple Observatory, Harvard-Smithsonian Center for Astrophysics, Amado, AZ 85645, USA }

\author{ T.~B.~Humensky }
\affiliation{ Physics Department, Columbia University, New York, NY 10027, USA }

\author{ C.~A.~Johnson }
\affiliation{ Santa Cruz Institute for Particle Physics and Department of Physics, University of California, Santa Cruz, CA 95064, USA }

\author{ P.~Kaaret }
\affiliation{ Department of Physics and Astronomy, University of Iowa, Van Allen Hall, Iowa City, IA 52242, USA }

\author{ P.~Kar }
\affiliation{ Department of Physics and Astronomy, University of Utah, Salt Lake City, UT 84112, USA }

\author{ N.~Kelley-Hoskins }
\affiliation{ DESY, Platanenallee 6, 15738 Zeuthen, Germany }

\author{ M.~Kertzman }
\affiliation{ Department of Physics and Astronomy, DePauw University, Greencastle, IN 46135-0037, USA }

\author{ D.~Kieda }
\affiliation{ Department of Physics and Astronomy, University of Utah, Salt Lake City, UT 84112, USA }

\author{ M.~Krause }
\affiliation{ DESY, Platanenallee 6, 15738 Zeuthen, Germany }

\author{ F.~Krennrich }
\affiliation{ Department of Physics and Astronomy, Iowa State University, Ames, IA 50011, USA }

\author{ M.~J.~Lang }
\affiliation{ School of Physics, National University of Ireland Galway, University Road, Galway, Ireland }

\author{ P.~Moriarty }
\affiliation{ School of Physics, National University of Ireland Galway, University Road, Galway, Ireland }

\author{ R.~Mukherjee }
\affiliation{ Department of Physics and Astronomy, Barnard College, Columbia University, NY 10027, USA }

\author{ S.~O'Brien }
\affiliation{ School of Physics, University College Dublin, Belfield, Dublin 4, Ireland }

\author{ R.~A.~Ong }
\affiliation{ Department of Physics and Astronomy, University of California, Los Angeles, CA 90095, USA }

\author{ A.~N.~Otte }
\affiliation{ School of Physics and Center for Relativistic Astrophysics, Georgia Institute of Technology, 837 State Street NW, Atlanta, GA 30332-0430 }

\author{ N.~Park }
\affiliation{ Wisconsin IceCube Particle Astrophysics Center, University of Wisconsin-Madison, Madison, WI 53703, USA }

\affiliation{ Enrico Fermi Institute, University of Chicago, Chicago, IL 60637, USA }

\author{ A.~Petrashyk }
\affiliation{ Physics Department, Columbia University, New York, NY 10027, USA }

\author{ M.~Pohl }
\affiliation{ Institute of Physics and Astronomy, University of Potsdam, 14476 Potsdam-Golm, Germany }
\affiliation{ DESY, Platanenallee 6, 15738 Zeuthen, Germany }

\author{ E.~Pueschel }
\affiliation{ DESY, Platanenallee 6, 15738 Zeuthen, Germany }

\author{ J.~Quinn }
\affiliation{ School of Physics, University College Dublin, Belfield, Dublin 4, Ireland }

\author{ K.~Ragan }
\affiliation{ Physics Department, McGill University, Montreal, QC H3A 2T8, Canada }

\author{ P.~T.~Reynolds }
\affiliation{ Department of Physical Sciences, Cork Institute of Technology, Bishopstown, Cork, Ireland }

\author{ G.~T.~Richards }
\affiliation{ School of Physics and Center for Relativistic Astrophysics, Georgia Institute of Technology, 837 State Street NW, Atlanta, GA 30332-0430 }

\author{ E.~Roache }
\affiliation{ Fred Lawrence Whipple Observatory, Harvard-Smithsonian Center for Astrophysics, Amado, AZ 85645, USA }

\author{ C.~Rulten }
\affiliation{ School of Physics and Astronomy, University of Minnesota, Minneapolis, MN 55455, USA }

\author{ I.~Sadeh }
\affiliation{ DESY, Platanenallee 6, 15738 Zeuthen, Germany }

\author{ M.~Santander }
\affiliation{ Department of Physics and Astronomy, University of Alabama, Tuscaloosa, AL 35487, USA }

\author{ S.~S.~Scott }
\affiliation{ Santa Cruz Institute for Particle Physics and Department of Physics, University of California, Santa Cruz, CA 95064, USA }

\author{ G.~H.~Sembroski }
\affiliation{ Department of Physics and Astronomy, Purdue University, West Lafayette, IN 47907, USA }

\author{ K.~Shahinyan }
\affiliation{ School of Physics and Astronomy, University of Minnesota, Minneapolis, MN 55455, USA }

\author{ J.~Tyler }
\affiliation{ Physics Department, McGill University, Montreal, QC H3A 2T8, Canada }

\author{ S.~P.~Wakely }
\affiliation{ Enrico Fermi Institute, University of Chicago, Chicago, IL 60637, USA }

\author{ A.~Weinstein }
\affiliation{ Department of Physics and Astronomy, Iowa State University, Ames, IA 50011, USA }

\author{ R.~M.~Wells }
\affiliation{ Department of Physics and Astronomy, Iowa State University, Ames, IA 50011, USA }

\author{ P.~Wilcox }
\affiliation{ Department of Physics and Astronomy, University of Iowa, Van Allen Hall, Iowa City, IA 52242, USA }

\author{ A.~Wilhelm }
\affiliation{ Institute of Physics and Astronomy, University of Potsdam, 14476 Potsdam-Golm, Germany }
\affiliation{ DESY, Platanenallee 6, 15738 Zeuthen, Germany }

\author{ D.~A.~Williams }
\affiliation{ Santa Cruz Institute for Particle Physics and Department of Physics, University of California, Santa Cruz, CA 95064, USA }

\author{ T.~J~Williamson }
\affiliation{ Department of Physics and Astronomy and the Bartol Research Institute, University of Delaware, Newark, DE 19716, USA }

\author{ B.~Zitzer }
\affiliation{ Physics Department, McGill University, Montreal, QC H3A 2T8, Canada }

\collaboration{(The VERITAS Collaboration)}
\noaffiliation

\author{ A.~Kaur }
\affiliation{ Department of Astronomy and Astrophysics, 525 Davey Lab, Pennsylvania State University, University Park, PA 16802, USA }

\correspondingauthor{Marcos Santander}
\email{jmsantander@ua.edu}



\begin{abstract}
On 2017 September 22, the IceCube Neutrino Observatory reported the detection of the high-energy neutrino event \icnu,  of potential astrophysical origin. It was soon determined that the neutrino direction was consistent with the location of the gamma-ray blazar \txs~(3FGL J0509.4+0541), which was in an elevated gamma-ray emission state as measured by the \emph{Fermi} satellite.
VERITAS observations of the neutrino/blazar region started on 2017 September 23 in response to the neutrino alert and continued through 2018 February 6. While no significant very-high-energy (VHE; E $>$ 100 GeV) emission was observed from the blazar by VERITAS in the two-week period immediately 
following the IceCube alert, TXS 0506+056 was detected by VERITAS with a significance of 5.8 standard deviations ($\sigma$) in the full 35-hour data set. The average photon flux of the source during this period was $(8.9 \pm 1.6) \times 10^{-12} \; \mathrm{cm}^{-2} \, \mathrm{s}^{-1}$, or 1.6\% of the Crab Nebula flux, above an energy threshold of 110 GeV, with a soft spectral index of $4.8 \pm 1.3$. 
\end{abstract}




\section{Introduction} \label{sec:intro}

The extragalactic gamma-ray sky is dominated by blazars~\citep{gammaagn, 2016ARA&A..54..725M}, a subclass of radio-loud active galactic nuclei (AGN) powered by a central supermassive black hole that displays relativistic jets, with one pointed close to the Earth's line of sight. 
The spectral energy distributions (SEDs) of blazars is characterized by two broad emission ``bumps''. The first one, in the radio to X-ray range, is believed to be due to synchrotron emission from relativistic electrons and positrons (henceforth electrons) in the jet. The origin of the second bump, in the X-ray to gamma-ray range, is less clear and is usually attributed in ``leptonic'' models to the inverse-Compton scattering of low-energy photons by electrons in the jet, and in ``hadronic'' models to proton synchrotron, or to the decay of high-energy mesons produced in cosmic ray interactions. See~\cite{2013ApJ...768...54B} and references therein for a recent summary of leptonic and hadronic modeling of blazar SEDs.

The potential hadronic origin of the high-energy emission from blazars~\citep{1993A&A...269...67M} makes them candidate neutrino sources~\citep{1995APh.....3..295M, 1997ApJ...488..669H} and, as such, they have been suggested as the origin of the astrophysical neutrino flux detected by the IceCube observatory at energies between $\sim20$ TeV and a few PeV~\citep{2013Sci...342E...1I}.
While the isotropic distribution of the IceCube astrophysical neutrinos favors an extragalactic origin, the source of the flux remains unknown and no significant correlation has been found in studies that involve neutrino positions and blazars detected by the \emph{Fermi}-LAT gamma-ray space telescope~\citep{2017ApJ...835...45A}, constraining the fractional contribution from \emph{Fermi}-LAT blazars to the all-sky astrophysical neutrino flux to be lower than 27\%. While the IceCube measurement excludes \emph{Fermi}-LAT blazars as the main source of the neutrino flux, the constraint has important caveats as it depends on the assumed neutrino flux spectrum, and the variability and other intrinsic characteristics of the sources considered. Therefore, it does not exclude the identification of an individual blazar as a potential neutrino counterpart.

In 2016 IceCube started broadcasting real-time alerts for astrophysical neutrino candidate events to  enable prompt follow-up observations that could identify electromagnetic counterparts to the astrophysical neutrinos. These alerts are issued by the Astrophysical Multimessenger Observatory Network (AMON)~\citep{2013APh....45...56S} and are circulated using the Gamma-ray Coordinates Network (GCN).\footnote{\url{https://gcn.gsfc.nasa.gov/}}

On 2017 September 22, IceCube reported the detection of a high-energy astrophysical neutrino candidate event, \icnu.\footnote{\url{https://gcn.gsfc.nasa.gov/notices_amon/50579430_130033.amon}} The initial reconstructed direction of the neutrino~\citep{2017GCN.21916....1K} was consistent with that of the gamma-ray blazar TXS 0506+056, which was observed to be in a high gamma-ray emission state by \emph{Fermi}-LAT~\citep{2017ATel10791....1T}. Follow-up observations of the neutrino/blazar position led to the first detection of this blazar in the VHE range by the MAGIC telescopes~\citep{2017ATel10817....1M}. VHE observations by HAWC~\citep{HAWCATel}, H.E.S.S.~\citep{HESSATel}, and an initial observation from VERITAS~\citep{VERITASATel} taken within two weeks of the event as part of its neutrino follow-up program~\citep{ICRC2017}, did not yield detections. Details of the neutrino detection, the multiwavelength follow-up campaign, and the statistical significance of the neutrino-blazar chance correlation are presented in~\cite{ScienceMM}. While potential counterparts to neutrino events have been proposed in the past~\citep{2017ApJ...846..121L,2016NatPh..12..807K}, none had previously been detected in the VHE range.

The potential association with an astrophysical neutrino event made TXS 0506+056 a source of particular interest, and further studies of its high-energy emission are required to explore the possible physical connection between the neutrino and gamma-ray emissions. For this purpose, VERITAS performed observations of the blazar between 2017 September and 2018 February which are presented here.

\section{Observations and Data Analysis} \label{sec:data_analysis}

\subsection{VERITAS Observations} \label{subsec:vts}

VERITAS ~\citep{Holder2006391} is an instrument dedicated to VHE gamma-ray astrophysics with sensitivity in the 80 GeV to 30 TeV energy range. It consists of an array of four 12-m imaging atmospheric Cherenkov telescopes (IACTs) located at the Fred Lawrence Whipple Observatory in southern Arizona, USA. 
In its current configuration, VERITAS is able to detect a source with a flux of 1\% of the gamma-ray flux of the Crab Nebula within 25 hours of observation~\citep{2015ICRC...34..771P}.

VERITAS observations of the TXS 0506+056 location were started on 2017 September 23 (MJD 58019.38) in response to the IceCube alert \icnu~and continued following the \emph{Fermi}-LAT report of flaring activity from the blazar and the detection in VHE gamma rays by the MAGIC collaboration. A total exposure of 34.9 hours of quality-selected data was accumulated between 2017 September 23 and 2018 February 6 (MJD 58019-58155) with an average zenith angle of $29^{\circ}$. Observations were performed using the standard ``wobble'' observation mode \citep{1994APh.....2..137F} with a $0.5^{\circ}$ offset in each of four cardinal directions. 

\begin{figure}[t]
\centering
\includegraphics[width=0.65\textwidth]{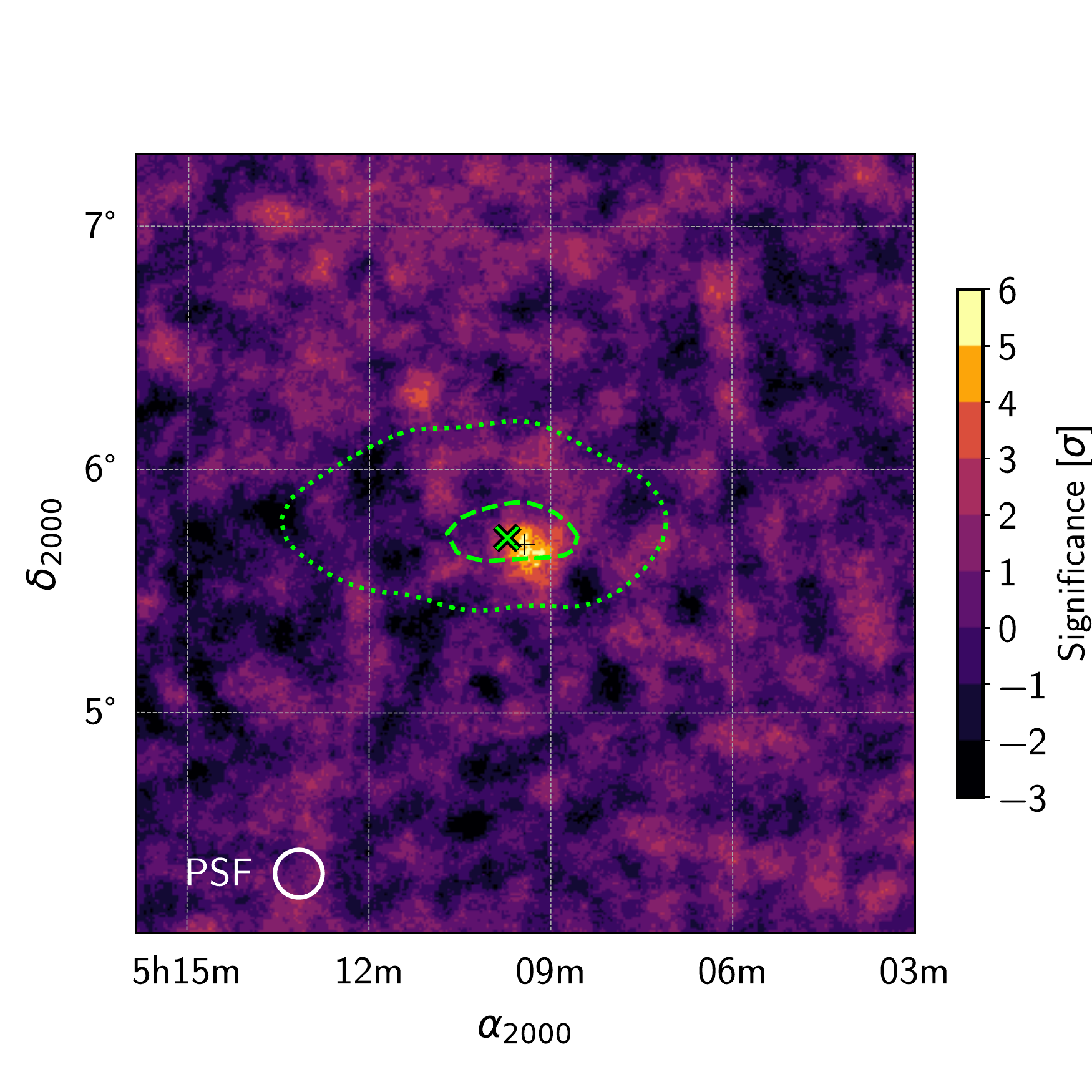}
\caption{VERITAS statistical-significance sky map for the region around TXS 0506+056. The VLBA radio location of the blazar is indicated with a `+' marker. The size of the VERITAS point spread function for this analysis, at 68\% containment, is shown as a white circle in the lower left. The `x' marker indicates the best-fit position of \icnu, with dashed (dotted) lines indicating the 50\% (90\%) confidence-level regions for the neutrino location (from~\cite{ScienceMM}).}
\label{fig:map}
\end{figure}

The data were analyzed using the standard VERITAS analysis tools \citep{2017arXiv170804048M,2008ICRC....3.1385C}, with background-rejection cuts optimized for soft-spectrum sources (photon index $\sim 4$, \cite{2017APh....89....1K}). 
The analysis yields a detection of the source above 110 GeV with a statistical significance of $5.8\sigma$ following Equation 17 of \cite{1983ApJ...272..317L}, with 1270 events in the ON region, 10647 events in the OFF region, and a background normalization $\alpha = 0.1$, resulting in a photon excess of 205 events in the ON region. A sky map of statistical significance is shown in Fig.~\ref{fig:map}. The centroid of the gamma-ray excess was determined by fitting the uncorrelated excess counts map with a two-dimensional Gaussian function. The centroid position in J2000 coordinates is RA=$77.354^\circ$ (or 5$^{\mathrm{h}}$ 9$^{\mathrm{m}}$ 25$^{\mathrm{s}}$), and Dec=$5.702^\circ$ (or 5$^{\circ}$ 42\arcmin 9\arcsec), in good agreement with the VLBA radio position of TXS 0506+056~\citep{2010AJ....139.1695L} given statistical and systematic uncertainties on each of the centroid coordinates of $0.01^{\circ}$ and $0.007^{\circ}$, respectively. We name the new VERITAS source VER J0509+057.

\begin{figure}[t]
\centering
\includegraphics[width=0.75\textwidth]{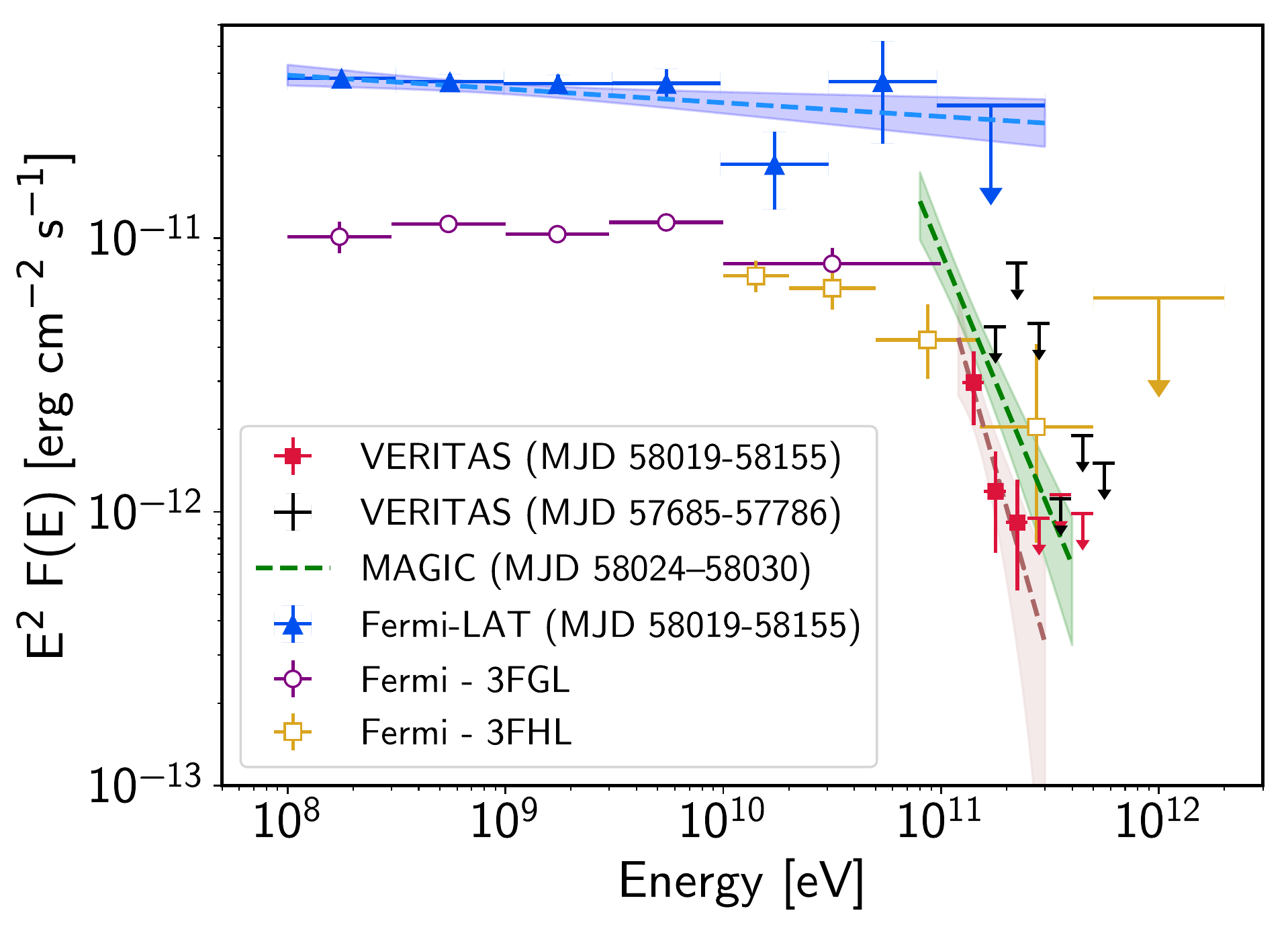}
\caption{Gamma-ray SED of TXS 0506+056 from \emph{Fermi}-LAT and VERITAS observations collected in the period MJD 58019-58155. Given that the observations are not strictly simultaneous, spectral variability of the source cannot be ruled out. The VHE spectrum of the source as measured by MAGIC within two weeks of the detection of \icnu~is also shown~\citep{ScienceMM}. For comparison, the \emph{Fermi}-LAT 3FGL (purple) and 3FHL (orange) catalog fluxes of the source are shown, as well as 95\% CL upper limits from VERITAS archival observations (black) described in Section~\ref{subsec:vts}. Best-fit power laws are shown as dashed lines for each data set collected in the MJD 58019-58155 period, with color bands indicating 68\% statistical uncertainties on the fit.}
\label{fig:sed}
\end{figure}

A differential photon spectrum, shown in Fig.~\ref{fig:sed}, was constructed using ten bins per energy decade in the VHE range. A power-law of the form $F(E) = N_{0} (E/E_0)^{-\Gamma}$ fit to the differential spectrum gives a best-fit flux normalization of $N_0 = (6.4 \pm 1.6) \times 10^{-11}$ cm$^{-2}$ s$^{-1}$ TeV$^{-1}$ at an energy of $E_0 = 0.15$ TeV and a spectral index $\Gamma = 4.8 \pm 1.3$, with a $\chi^2/$n.d.f. of 0.5/1. The integral flux above 110 GeV is $(8.9 \pm 1.6) \times 10^{-12}$ cm$^{-2}$ s$^{-1}$, which corresponds to 1.6\% of the Crab Nebula flux~\citep{1998ApJ...503..744H} above the same energy threshold. The average spectrum measured by VERITAS between 2017 September and 2018 February has a flux of $\sim60$\% of that reported by MAGIC in observations obtained within two weeks of the \icnu~  detection~\citep{ScienceMM}. The spectral indices are consistent given the large statistical uncertainties ($\Gamma_{\mathrm{VERITAS}} = 4.8 \pm 1.3$ and $\Gamma_{\mathrm{MAGIC}} = 3.9 \pm 0.4$). Monte-Carlo simulations indicate that the systematic uncertainty on the VERITAS flux normalization and photon index for a soft-spectrum source like TXS 0506+056 are $\sim$60\% and 0.3, respectively~\citep{2015ApJ...815L..22A}.

A VERITAS light curve of the integral photon flux above an energy threshold of 110 GeV  is shown in Fig.\ref{fig:lc}. For its calculation, the spectral index has been kept fixed at 4.8.  Both in the case of the light curve and the differential spectrum, flux points are shown if the significance of the photon excess in the bin is larger than $2\sigma$; otherwise a 95\% confidence level (CL) upper limit is calculated using the method of \cite{2001NIMPA.458..745R}. 

Due to its hard GeV-band spectrum, TXS 0506+056 was identified as a promising VHE source candidate and observed by VERITAS prior to the detection of \icnu, between 2016 October 10 (MJD 57685) and 2017 February 2 (MJD 57786). No significant gamma-ray excess was found in an analysis of 2.1 hours of quality-selected data collected on the source during this period. A 95\% CL integral flux upper limit of 1.2 $\times 10^{-11}$ cm$^{-2}$ s$^{-1}$ above an energy threshold of 110 GeV was derived at the source location for a spectral index of 4.8, corresponding to 2.1\% of the Crab Nebula flux above the same threshold. Differential spectral upper limits computed at the 95\% CL are shown in Fig.~\ref{fig:sed} which, given the short exposure, are consistent with the VHE spectrum of the source measured after \icnu.  

\subsection{Fermi-LAT Observations} \label{subsec:fermi}

In order to characterize the GeV energy spectrum of the source during the period of VERITAS observations, we perform a power-law fit to data from the \emph{Fermi} LAT~\citep{2009ApJ...697.1071A} recorded between MJD 58019.38 and 58155.20. The analysis presented here was performed using version \texttt{v10r0p5} of the \emph{Fermi} \texttt{Science Tools}.\footnote{The \emph{Fermi} \texttt{Science Tools} can be downloaded from \url{https://fermi.gsfc.nasa.gov/ssc/data/analysis/software/}} 
Photons with energies between 100 MeV and 300 GeV that were detected within $15^{\circ}$ of the location of TXS 0506+056 were selected for the analysis, while photons with a zenith angle larger than $100^{\circ}$ were discarded to reduce contamination from the Earth's albedo.  
The contribution from isotropic and Galactic diffuse backgrounds, and sources in the 3FGL catalog~\citep{3FGL} within $15^{\circ}$ of the source position, were included in the spectral fit with their spectral parameters fixed to their catalog values, while the parameters for sources within $3^{\circ}$ were allowed to vary freely during the source spectral fit. The blazar spectral fit was performed with a binned-likelihood method using the \texttt{P8R2\_SOURCE\_V6} instrument response functions. 

TXS 0506+056 is strongly detected during the analyzed period, with a test-statistic (TS) of more than 2100 from the \emph{Fermi}-LAT analysis. The power-law best-fit spectral parameters are a photon index $\Gamma = 2.05 \pm 0.03$ (consistent with the 3FGL value of $2.04 \pm 0.03$) and a flux normalization $N_0 = (1.04 \pm 0.05) \times 10^{-11}$ cm$^{-2}$ s$^{-1}$ MeV$^{-1}$ at an energy $E_0$ of 1.44 GeV, about a factor of three higher than the 3FGL value of $(3.24 \pm 0.10) \times 10^{-12}$ in the same units. The spectral fit was repeated in seven independent energy bins with equal logarithmic spacing in the 0.1 - 300 GeV range. Best-fit flux values and 68\% uncertainties, shown in  Fig.~\ref{fig:sed}, are reported for spectral bins with a TS larger than 4. 
Flux upper limits at 95\% CL are quoted otherwise. 

\begin{figure}[t]
\centering
\includegraphics[width=\textwidth]{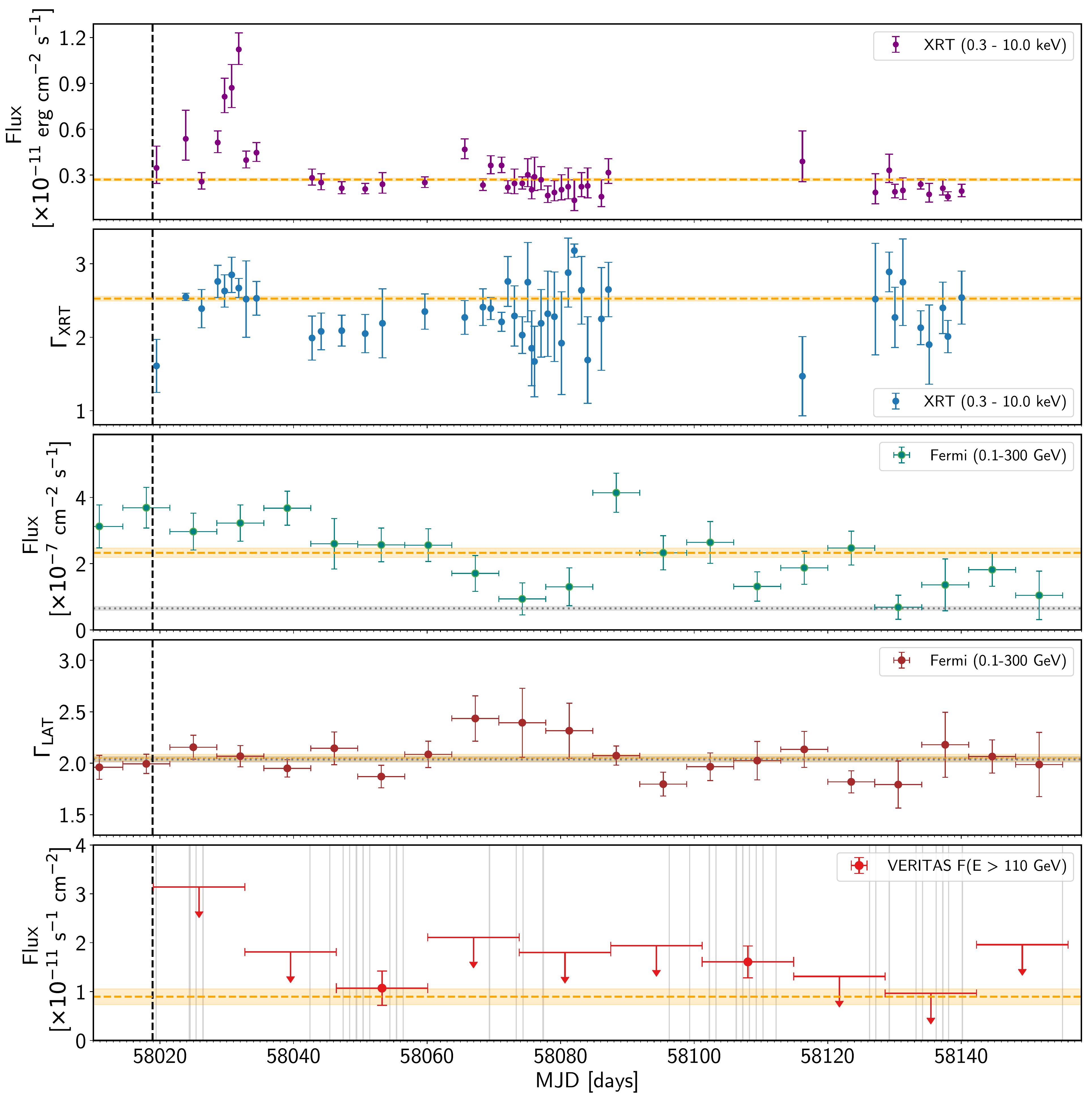}
\caption{Multiwavelength light curves for TXS 0506+056 derived from \emph{Swift}, \emph{Fermi}-LAT and VERITAS observations. The dashed vertical line shows the time of detection of the \icnu~neutrino event. In each panel, the horizontal dashed line and band indicate the mean and $1\sigma$ uncertainty of each parameter for the entire period. The catalog flux and spectral index values, and their uncertainties, are shown as gray lines and bands in the \emph{Fermi}-LAT panels. The VERITAS light curve is shown using ten bins, where each point represents the average flux level of the source during the VERITAS observations collected in that period (shown as vertical gray bands). The upper limits and flux points do not constrain the potential variability of the source during periods in which VERITAS was not observing the blazar.}
\label{fig:lc}
\end{figure}

We compute a light curve of integral photon flux between 0.1 and 300 GeV based on the \emph{Fermi}-LAT data binned in seven-day intervals. The integral flux is calculated by fitting the spectrum of the source with a power-law function for each time bin while allowing the spectral index $\Gamma_{\mathrm{LAT}}$ to vary freely. The integral flux points, all with a TS larger than 9, are shown in Fig.~\ref{fig:lc}. For comparison, the integral photon flux in the same energy range is $(2.33 \pm 0.14) \times 10^{-7}$ cm$^{-2}$ s$^{-1}$ for the entire period. We characterize the variability of the source in the \emph{Fermi}-LAT band by evaluating the goodness-of-fit of a constant value to the light curve. The flux light curve is poorly fit by a constant, with a $\chi^2$/n.d.f. of 69.4/19 (\emph{p} = 1.2 $\times 10^{-7}$), which confirms the presence of GeV flux variability in the time period considered. For the spectral index, a constant fit yields a $\chi^2$/n.d.f. of 23.9/19 (\emph{p} = 0.2), indicating that a constant index is consistent with the data given the statistical uncertainties. 

\subsection{Swift-XRT Observations} \label{subsec:xrt}

Prompt follow-up observations of the neutrino error region and later observations of TXS 0506+056 were performed as part of an existing neutrino follow-up program~\citep{2015MNRAS.448.2210E} using the X-ray Telescope (XRT) onboard the Neil Gehrels \emph{Swift} Observatory~\citep{2004AIPC..727..637G}. Preliminary results from this follow-up were reported by \cite{2017ATel10792....1E}. A total of 45 \emph{Swift} XRT observations performed during the MJD 58019-58155 period resulted in a total exposure of 74 ks, which we use to characterize the X-ray variability of the source. The observations were conducted in photon-counting mode, with negligible pile-up effects. The data were reduced and calibrated using HEAsoft, XSpec version 12.9.1. The data from each observation (typically $\sim1$ ksec) were fitted with an absorbed power-law spectrum using a Galactic column density of $1.1 \times 10^{21}$ cm$^{-2}$ from the LAB neutral hydrogen survey~\citep{LAB} and using \texttt{cstat} within XSpec. A good fit was obtained for all observations considered.

The de-absorbed integral photon fluxes and indices in the 0.3-10 keV band from the individual fits are shown in the light curve in Fig.~\ref{fig:lc}. The flux is clearly variable during the period both in amplitude and spectral index, with a strong X-ray flare detected in the period MJD 58030-58032 peaking at a flux about four times the time-average for the period, which is $(2.7 \pm 0.1) \times 10^{-12}$ erg cm$^{-2}$ s$^{-2}$. The time-averaged photon index in XRT was $2.52 \pm 0.03$. 

As in the \emph{Fermi}-LAT case, a quantitative assessment of the flux variability was performed by fitting a constant value to the flux and index curves, both resulting in poor fits ($\chi^2$/ndf of 213/44 and 118/44, respectively) indicating source variability.

\section{Discussion}

Our analysis indicates that during this period the blazar was about three times brighter in the LAT band compared to its 3FGL value and, while there was significant GeV photon flux variability, the photon index remained consistent with a constant value at the 3FGL level. At the same time, the flux observed by VERITAS is consistent with that reported in a similar energy range in the \emph{Fermi}-LAT 3FHL catalog~\citep{3FHL}. This could imply that the VERITAS detection is associated with a baseline VHE flux of the source rather than with flaring episodes. As the VERITAS light curve upper limits are consistent with the average flux for the entire period, this hypothesis cannot be ruled out using these data alone. While the variability detected in the LAT and XRT bands has no clear counterpart in the VHE band, the upper limits set by VERITAS are still consistent with a linear scaling of the VHE flux by the same amount as the flux increase observed at lower energies during the active periods, with the additional caveat that the VERITAS observations are not strictly simultaneous with those of XRT and LAT. 

The gamma-ray SED in Fig.~\ref{fig:sed} shows a sharp spectral break between the \emph{Fermi}-LAT and VERITAS energy bands during the MJD 58019-58155 period. Although not as pronounced, some spectral curvature is also evident in the flux points reported in the 3FGL and 3FHL catalogs shown in the same figure. This softening is noticeable in the power-law indices reported in \emph{Fermi}-LAT catalogs, which increase with energy threshold ($\Gamma_{\mathrm{3FGL}} = 2.04 \pm 0.03$ for $E > 100$ MeV, $\Gamma_{\mathrm{3FHL}} = 2.16 \pm 0.21$ for $E > 10$ GeV, and $\Gamma_{\mathrm{2FHL}} = 2.76 \pm 0.57$ for $E > 50$ GeV), although with significant uncertainties. 

\begin{figure}[t]
\centering
\includegraphics[width=0.75\textwidth]{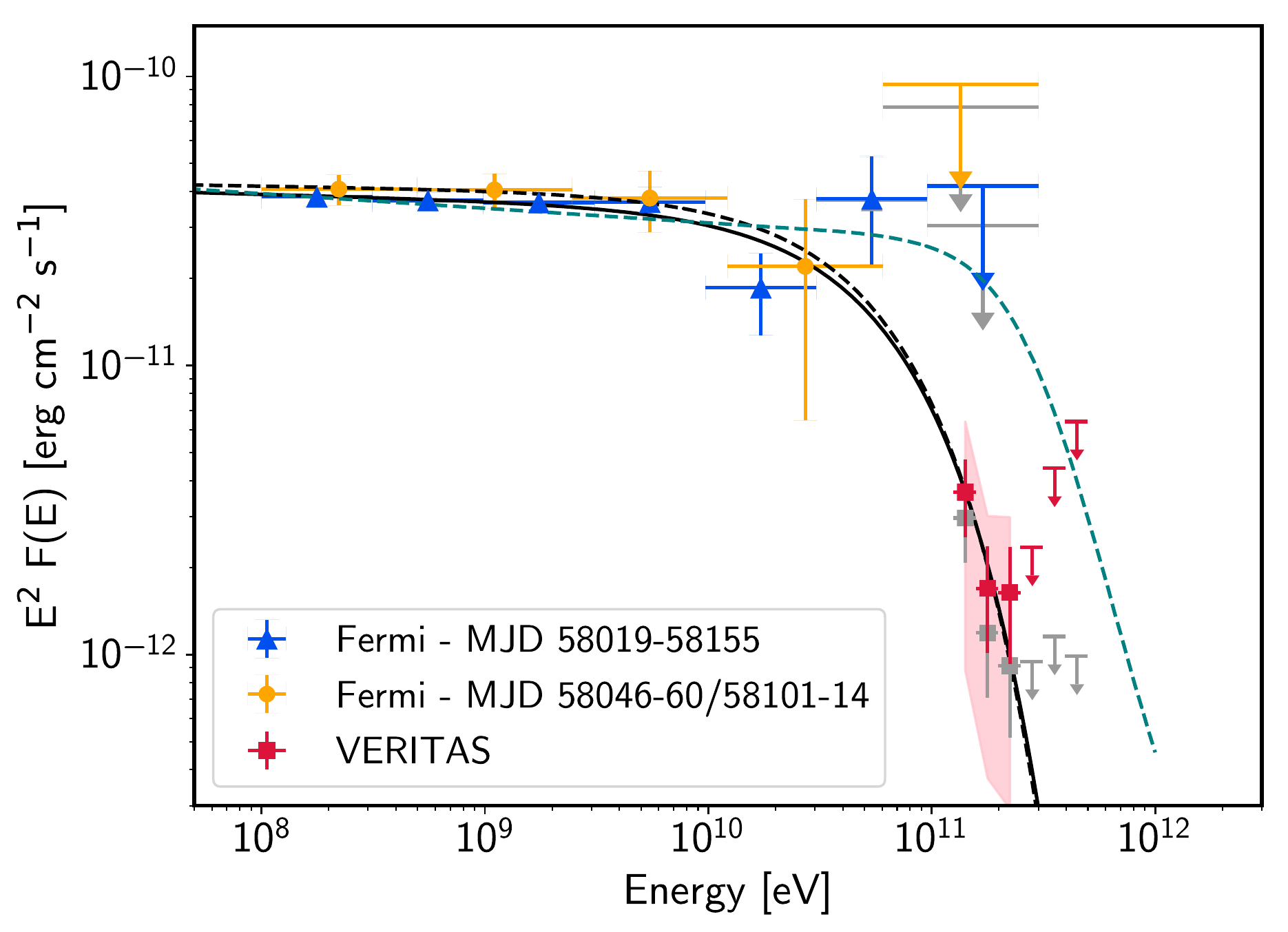}
\caption{Power-law fit with an exponential cutoff to the de-absorbed gamma-ray SED of TXS 0506+056 using \emph{Fermi}-LAT data from the period MJD 58019-58155 (solid black line) and MJD 58046-60/58101-58114 (dashed black line).
De-absorbed flux points are shown in color, and observed fluxes in gray. The teal dashed line shows an extrapolation of the \emph{Fermi}-LAT fit from Subsection~\ref{subsec:fermi} to VHE energies that accounts for EBL absorption using the model of \cite{2008A&A...487..837F}. The pink band in the VHE range illustrates the energy-scale systematic uncertainty on the VERITAS spectrum.}
\label{fig:sedabs}
\end{figure}

To characterize the intrinsic break we correct the VERITAS and \emph{Fermi}-LAT flux points for extragalactic background light (EBL) absorption using the model of \cite{2008A&A...487..837F} and the source redshift of $z = 0.3365 \pm 0.0010$ ~\citep{TXSRedshift} and perform a joint fit of the gamma-ray spectrum. 
The SED is well-described using a power-law function with exponential cutoff of the form $N_0 (E/E_0) ^{-\Gamma} \exp(-E/E_c)$. For a normalization energy $E_0$ of 1.44 GeV the best-fit parameters are $N_0 = (1.12 \pm 0.05) \times 10^{-8}$  cm$^{-2}$ s$^{-1}$ GeV$^{-1}$, $E_c = 63 \pm 7$ GeV, and $\Gamma = 2.02 \pm 0.03$, with a $\chi^2$/n.d.f. of $6.05/6$ ($p$ = 0.42).
The de-absorbed spectrum and best-fit solution are shown in Fig.\ref{fig:sedabs}. 
This result indicates that the break cannot be attributed to EBL absorption of a power-law spectrum alone, as illustrated by the extrapolation of the \emph{Fermi}-LAT power-law spectrum to VHE energies which fails to fit the VERITAS observations after accounting for EBL absorption. If the blazar is the source of the neutrino, the paucity of high-energy emission may imply that the hadronic gamma-ray emission is strongly attenuated at the source, and potentially cascades down to energies lower than the \emph{Fermi}-LAT band~\citep{2016PhRvL.116g1101M, 2017ApJ...843..109G}. The soft XRT photon index and lower XRT energy flux compared to the $>100$ MeV band places the X-ray emission in the first SED bump, pushing any potential cascaded emission to the hard X-ray to MeV band. Future measurements of the ``valley'' between the SED bumps will be crucial in constraining the contribution to the SED of potential neutrino counterparts from cascaded emission.

The sharpness of the break may be sensitive to relative flux variations between the \emph{Fermi}-LAT and VERITAS bands as the observations are not strictly simultaneous. As shown in Fig.~\ref{fig:lc}, most of the VERITAS signal is collected during two time periods: MJD 58046-60 and 58101-58114. We obtained a second \emph{Fermi}-LAT spectrum, also shown in Fig.\ref{fig:sedabs}, by restricting the analysis to this period and repeated the SED fit which yields best-fit parameters of $N_0 = (1.2 \pm 0.1) \times 10^{-8}$  cm$^{-2}$ s$^{-1}$ GeV$^{-1}$, $E_c = 61 \pm 11$ GeV, and $\Gamma = 2.01 \pm 0.07$, with a $\chi^2$/n.d.f. of $1.3/4$ ($p$ = 0.86), consistent with the fit for the entire period. This indicates that the break between both bands is robust to the observed relative flux variability, even after accounting for systematic uncertainties in the VERITAS energy scale (Fig.~\ref{fig:sedabs}).

The detection of TXS 0506+056 in the VHE band during extended follow-up observations of the \icnu~event represents the first time that such observations have revealed a new VHE gamma-ray source. This detection will inform future neutrino follow-up strategies, as the potential gamma-ray counterparts may be active over time periods of weeks or months, requiring multiple exposures. Continued observations in the future could clarify if blazars are indeed sources of astrophysical neutrinos detected by IceCube. 

\acknowledgments

This research is supported by grants from the U.S. Department of Energy Office of Science, the U.S. National Science Foundation and the Smithsonian Institution, and by NSERC in Canada. We acknowledge the excellent work of the technical support staff at the Fred Lawrence Whipple Observatory and at the collaborating institutions in the construction and operation of the instrument.

This research has made use of publicly available
\emph{Fermi}-LAT data and analysis software obtained from the \emph{Fermi} Science Support
Center at the High Energy Astrophysics Science Archive Research Center
(HEASARC), provided by NASA's Goddard Space Flight Center.
This research made use of Astropy, a community-developed core Python package for Astronomy \citep{2013A&A...558A..33A}; matplotlib, a Python library for publication quality graphics \citep{Hunter:2007}; and NumPy \citep{van2011numpy}.

%

\vspace{5mm}
\facilities{VERITAS, \emph{Fermi}-LAT, \emph{Swift}}

\bibliographystyle{aasjournal}

\begin{thebibliography}{}
\expandafter\ifx\csname natexlab\endcsname\relax\def\natexlab#1{#1}\fi
\providecommand{\url}[1]{\href{#1}{#1}}

\bibitem[{{Aartsen} {et~al.}(2017){Aartsen}, {Abraham}, {Ackermann},
  {et~al.}}]{2017ApJ...835...45A}
{Aartsen}, M.~G., {Abraham}, K., {Ackermann}, M., {et~al.} 2017, \apj, 835, 45

\bibitem[{{Abeysekara} {et~al.}(2015){Abeysekara}, {Archambault}, {Archer},
  {et~al.}}]{2015ApJ...815L..22A}
{Abeysekara}, A.~U., {Archambault}, S., {Archer}, A., {et~al.} 2015, \apjl,
  815, L22

\bibitem[{{Acero} {et~al.}(2015){Acero}, {Ackermann}, {Ajello},
  {et~al.}}]{3FGL}
{Acero}, F., {Ackermann}, M., {Ajello}, M., {et~al.} 2015, \apjs, 218, 23

\bibitem[{{Ajello} {et~al.}(2017){Ajello}, {Atwood}, {Baldini},
  {et~al.}}]{3FHL}
{Ajello}, M., {Atwood}, W.~B., {Baldini}, L., {et~al.} 2017, \apjs, 232, 18

\bibitem[{{Astropy Collaboration} {et~al.}(2013){Astropy Collaboration},
  {Robitaille}, {Tollerud}, {Greenfield}, {Droettboom}, {Bray}, {Aldcroft},
  {Davis}, {Ginsburg}, {Price-Whelan}, {Kerzendorf}, {Conley}, {Crighton},
  {Barbary}, {Muna}, {Ferguson}, {Grollier}, {Parikh}, {Nair}, {Unther},
  {Deil}, {Woillez}, {Conseil}, {Kramer}, {Turner}, {Singer}, {Fox}, {Weaver},
  {Zabalza}, {Edwards}, {Azalee Bostroem}, {Burke}, {Casey}, {Crawford},
  {Dencheva}, {Ely}, {Jenness}, {Labrie}, {Lim}, {Pierfederici}, {Pontzen},
  {Ptak}, {Refsdal}, {Servillat}, \& {Streicher}}]{2013A&A...558A..33A}
{Astropy Collaboration}, {Robitaille}, T.~P., {Tollerud}, E.~J., {et~al.} 2013,
  \aap, 558, A33

\bibitem[{{Atwood} {et~al.}(2009){Atwood}, {Abdo}, {Ackermann},
  {et~al.}}]{2009ApJ...697.1071A}
{Atwood}, W.~B., {Abdo}, A.~A., {Ackermann}, M., {et~al.} 2009, \apj, 697, 1071

\bibitem[{{B{\"o}ttcher} {et~al.}(2013){B{\"o}ttcher}, {Reimer}, {Sweeney}, \&
  {Prakash}}]{2013ApJ...768...54B}
{B{\"o}ttcher}, M., {Reimer}, A., {Sweeney}, K., \& {Prakash}, A. 2013, \apj,
  768, 54

\bibitem[{{Cogan}(2007)}]{2008ICRC....3.1385C}
{Cogan}, P. 2007, Proc. of 30th ICRC, Vol 3, 1385-1388, 3, 1385

\bibitem[{{de Naurois}(2017)}]{HESSATel}
{de Naurois}, M. 2017, The Astronomer's Telegram, 10787

\bibitem[{{Dermer} \& {Giebels}(2016)}]{gammaagn}
{Dermer}, C.~D., \& {Giebels}, B. 2016, C R Phys., 17, 594

\bibitem[{{Evans} {et~al.}(2015){Evans}, {Osborne}, {Kennea},
  {et~al.}}]{2015MNRAS.448.2210E}
{Evans}, P.~A., {Osborne}, J.~P., {Kennea}, J.~A., {et~al.} 2015, \mnras, 448,
  2210

\bibitem[{{Fomin} {et~al.}(1994){Fomin}, {Stepanian}, {Lamb}, {Lewis}, {Punch},
  \& {Weekes}}]{1994APh.....2..137F}
{Fomin}, V.~P., {Stepanian}, A.~A., {Lamb}, R.~C., {et~al.} 1994, Astroparticle
  Physics, 2, 137

\bibitem[{{Franceschini} {et~al.}(2008){Franceschini}, {Rodighiero}, \&
  {Vaccari}}]{2008A&A...487..837F}
{Franceschini}, A., {Rodighiero}, G., \& {Vaccari}, M. 2008, \aap, 487, 837

\bibitem[{{Gao} {et~al.}(2017){Gao}, {Pohl}, \& {Winter}}]{2017ApJ...843..109G}
{Gao}, S., {Pohl}, M., \& {Winter}, W. 2017, \apj, 843, 109

\bibitem[{{Gehrels}(2004)}]{2004AIPC..727..637G}
{Gehrels}, N. 2004, in AIP Conf Series, 637-641, Vol. 727, 637--641

\bibitem[{{Halzen} \& {Zas}(1997)}]{1997ApJ...488..669H}
{Halzen}, F., \& {Zas}, E. 1997, \apj, 488, 669

\bibitem[{{Hillas} {et~al.}(1998){Hillas}, {Akerlof}, {Biller},
  {et~al.}}]{1998ApJ...503..744H}
{Hillas}, A.~M., {Akerlof}, C.~W., {Biller}, S.~D., {et~al.} 1998, \apj, 503,
  744

\bibitem[{Holder {et~al.}(2006)Holder, Atkins, Badran,
  {et~al.}}]{Holder2006391}
Holder, J., Atkins, R., Badran, H., {et~al.} 2006, Astroparticle Physics, 25,
  391

\bibitem[{Hunter(2007)}]{Hunter:2007}
Hunter, J.~D. 2007, Computing In Science \& Engineering, 9, 90

\bibitem[{{IceCube Collaboration}(2013)}]{2013Sci...342E...1I}
{IceCube Collaboration}. 2013, Science, 342, 1242856

\bibitem[{{IceCube Collaboration} {et~al.}(2018)}]{ScienceMM}
{IceCube Collaboration}, {et~al.} 2018, Science (submitted)

\bibitem[{{Kadler} {et~al.}(2016){Kadler}, {Krau{\ss}}, {Mannheim}, {Ojha},
  {M{\"u}ller}, {Schulz}, {Anton}, {Baumgartner}, {Beuchert}, {Buson},
  {Carpenter}, {Eberl}, {Edwards}, {Eisenacher Glawion}, {Els{\"a}sser},
  {Gehrels}, {Gr{\"a}fe}, {Gulyaev}, {Hase}, {Horiuchi}, {James}, {Kappes},
  {Kappes}, {Katz}, {Kreikenbohm}, {Kreter}, {Kreykenbohm}, {Langejahn},
  {Leiter}, {Litzinger}, {Longo}, {Lovell}, {McEnery}, {Natusch}, {Phillips},
  {Pl{\"o}tz}, {Quick}, {Ros}, {Stecker}, {Steinbring}, {Stevens}, {Thompson},
  {Tr{\"u}stedt}, {Tzioumis}, {Weston}, {Wilms}, \&
  {Zensus}}]{2016NatPh..12..807K}
{Kadler}, M., {Krau{\ss}}, F., {Mannheim}, K., {et~al.} 2016, Nature Physics,
  12, 807

\bibitem[{{Kalberla} {et~al.}(2005){Kalberla}, {Burton}, {Hartmann},
  {et~al.}}]{LAB}
{Kalberla}, P.~M.~W., {Burton}, W.~B., {Hartmann}, D., {et~al.} 2005, \aap,
  440, 775

\bibitem[{{Keivani} {et~al.}(2017){Keivani}, {Evans}, {Kennea},
  {et~al.}}]{2017ATel10792....1E}
{Keivani}, A., {Evans}, P.~A., {Kennea}, J.~A., {et~al.} 2017, The Astronomer's
  Telegram, 10792

\bibitem[{{Kopper} \& {Blaufuss}(2017)}]{2017GCN.21916....1K}
{Kopper}, C., \& {Blaufuss}, E. 2017, GRB Coordinates Network, Circular
  Service, No.~21916, \#1 (2017), 21916

\bibitem[{{Krause} {et~al.}(2017){Krause}, {Pueschel}, \&
  {Maier}}]{2017APh....89....1K}
{Krause}, M., {Pueschel}, E., \& {Maier}, G. 2017, Astroparticle Physics, 89, 1

\bibitem[{{Lanyi} {et~al.}(2010){Lanyi}, {Boboltz}, {Charlot},
  {et~al.}}]{2010AJ....139.1695L}
{Lanyi}, G.~E., {Boboltz}, D.~A., {Charlot}, P., {et~al.} 2010, \aj, 139, 1695

\bibitem[{{Li} \& {Ma}(1983)}]{1983ApJ...272..317L}
{Li}, T.-P., \& {Ma}, Y.-Q. 1983, \apj, 272, 317

\bibitem[{{Lucarelli} {et~al.}(2017){Lucarelli}, {Pittori}, {Verrecchia},
  {Donnarumma}, {Tavani}, {Bulgarelli}, {Giuliani}, {Antonelli}, {Caraveo},
  {Cattaneo}, {Colafrancesco}, {Longo}, {Mereghetti}, {Morselli}, {Pacciani},
  {Piano}, {Pellizzoni}, {Pilia}, {Rappoldi}, {Trois}, \&
  {Vercellone}}]{2017ApJ...846..121L}
{Lucarelli}, F., {Pittori}, C., {Verrecchia}, F., {et~al.} 2017, \apj, 846, 121

\bibitem[{{Madejski} \& {Sikora}(2016)}]{2016ARA&A..54..725M}
{Madejski}, G., \& {Sikora}, M. 2016, \araa, 54, 725

\bibitem[{{Maier} \& {Holder}(2017)}]{2017arXiv170804048M}
{Maier}, G., \& {Holder}, J. 2017, ArXiv e-prints, arXiv:1708.04048

\bibitem[{{Mannheim}(1993)}]{1993A&A...269...67M}
{Mannheim}, K. 1993, \aap, 269, 67

\bibitem[{{Mannheim}(1995)}]{1995APh.....3..295M}
---. 1995, Astroparticle Physics, 3, 295

\bibitem[{{Martinez} {et~al.}(2017){Martinez}, {Taboada}, {Hui}, \&
  {Lauer}}]{HAWCATel}
{Martinez}, I., {Taboada}, I., {Hui}, M., \& {Lauer}, R. 2017, The Astronomer's
  Telegram, 10802

\bibitem[{{Mirzoyan}(2017)}]{2017ATel10817....1M}
{Mirzoyan}, R. 2017, The Astronomer's Telegram, 10817

\bibitem[{{Mukherjee}(2017)}]{VERITASATel}
{Mukherjee}, R. 2017, The Astronomer's Telegram, 10833

\bibitem[{{Murase} {et~al.}(2016){Murase}, {Guetta}, \&
  {Ahlers}}]{2016PhRvL.116g1101M}
{Murase}, K., {Guetta}, D., \& {Ahlers}, M. 2016, Physical Review Letters, 116,
  071101

\bibitem[{{Paiano} {et~al.}(2018){Paiano}, {Falomo}, {Treves}, \&
  {Scarpa}}]{TXSRedshift}
{Paiano}, S., {Falomo}, R., {Treves}, A., \& {Scarpa}, R. 2018, \apjl, 854, L32

\bibitem[{{Park} {et~al.}(2015)}]{2015ICRC...34..771P}
{Park}, N., {et~al.} 2015, Proc. of 34th ICRC, Vol 34, 771, 34, 771

\bibitem[{{Rolke} \& {L{\'o}pez}(2001)}]{2001NIMPA.458..745R}
{Rolke}, W.~A., \& {L{\'o}pez}, A.~M. 2001, Nuclear Instruments and Methods in
  Physics Research A, 458, 745

\bibitem[{{Santander} {et~al.}(2017){Santander}, {Dorner}, {Dumm},
  {et~al.}}]{ICRC2017}
{Santander}, M., {Dorner}, D., {Dumm}, J., {et~al.} 2017, ArXiv e-prints,
  arXiv:1708.08945

\bibitem[{{Smith} {et~al.}(2013){Smith}, {Fox}, {Cowen},
  {et~al.}}]{2013APh....45...56S}
{Smith}, M.~W.~E., {Fox}, D.~B., {Cowen}, D.~F., {et~al.} 2013, Astroparticle
  Physics, 45, 56

\bibitem[{{Tanaka} {et~al.}(2017){Tanaka}, {Buson}, \&
  {Kocevski}}]{2017ATel10791....1T}
{Tanaka}, Y.~T., {Buson}, S., \& {Kocevski}, D. 2017, The Astronomer's
  Telegram, 10791

\bibitem[{Van Der~Walt {et~al.}(2011)Van Der~Walt, Colbert, \&
  Varoquaux}]{van2011numpy}
Van Der~Walt, S., Colbert, S.~C., \& Varoquaux, G. 2011, Computing in Science
  \& Engineering, 13, 22

\end{thebibliography}






\end{document}